\documentclass[prl,twocolumn,groupedaddress,nofootinbib]{revtex4}

\usepackage{amsmath,amssymb,amscd}
\usepackage{listings}
\usepackage{dsfont}
\usepackage{slashed}
\usepackage{color}
\usepackage{ulem}

\usepackage[pdftex]{graphicx}
\usepackage{epstopdf}
\usepackage{subfigure}
\usepackage{epsfig}
\begin{document}






\title{Light-by-Light Scattering Constraint on Born-Infeld Theory}

%
%
%
%
%
%
%
%
%
%

\author{John~Ellis}
 
 \affiliation{Theoretical Particle Physics and Cosmology Group, Physics Department, 
King's College London, London WC2R 2LS, UK ;\\
Theoretical Physics Department, CERN, CH-1211 Geneva 23, Switzerland}

\author{Nick~E.~Mavromatos}

 \affiliation{Theoretical Particle Physics and Cosmology Group, Physics Department, 
King's College London, London WC2R 2LS, UK}

\author{Tevong~You}

\affiliation{DAMTP, University of Cambridge, Wilberforce Road, Cambridge, CB3 0WA, UK; \\
Cavendish Laboratory, University of Cambridge, J.J. Thomson Avenue, Cambridge, CB3 0HE, UK}

\begin{abstract}
The recent measurement by ATLAS of light-by-light scattering in LHC Pb-Pb collisions is the first direct evidence for this basic process. We find that it excludes
a range of the mass scale of 
a nonlinear Born-Infeld extension of QED that is $\lesssim 100$~GeV,  a much stronger constraint than those derived previously. In the case of a Born-Infeld extension of the 
Standard Model in which the U(1)$_{\rm Y}$ hypercharge gauge symmetry is realized nonlinearly, the limit on the corresponding mass reach is $\sim 90$~GeV, 
which in turn imposes a lower limit of $\gtrsim 11$~TeV on the magnetic monopole mass in such a U(1)$_{\rm Y}$ Born-Infeld theory.
\end{abstract}

\maketitle 

%
%
%
%


Over 80 years ago, soon after Dirac proposed his relativistic theory of the electron~\cite{Dirac28}
and his interpretation of `hole' states as positrons~\cite{Dirac31}, Halpern~\cite{Halpern33} in 1933 and Heisenberg~\cite{Heisenberg34} in 1934 realized that
quantum effects would induce light-by-light scattering, which was first calculated in the low-frequency limit by Euler and Kockel~\cite{EulerKockel35} in 1935.
Subsequently, Heisenberg and Euler~\cite{HeisenbergEuler36} derived in 1936 a more general expression for the quantum nonlinearities in the Lagrangian of Quantum Electrodynamics (QED),
and a complete calculation of light-by-light scattering in QED was published by Karplus and Neuman~\cite{KarplusNeuman51} in 1951.
However, measurement of light-by-light scattering has remained elusive until very recently. In 2013 d'Enterria and Silveira~\cite{dES13} proposed looking for
light-by-light scattering in ultraperipheral heavy-ion collisions at the LHC, and evidence for this process was recently presented by the ATLAS Collaboration~\cite{ATLAS17},
at a level consistent with the QED predictions in~\cite{dES13} and~\cite{K-GLS16}.

In parallel with the early work on light-by-light scattering in QED, and motivated by a `unitarian' idea that there should be an upper
limit on the strength of the electromagnetic field, Born and Infeld~\cite{BornInfeld34} proposed in 1934
a conceptually distinct nonlinear modification of the Lagrangian of QED:
\begin{eqnarray}
&& {\cal L}_{\rm QED} \; = \; - \frac{1}{4} F_{\mu \nu} F^{\mu \nu} \: \to \;  \nonumber \\ 
&& {\cal L}_{\rm BI} \; = \; \beta^2 \Big(1
 - \sqrt{1 + \frac{1}{2 \beta^2} F_{\mu \nu} F^{\mu \nu}
-\frac{1}{16  \beta^4} (F_{\mu \nu} \tilde{F}^{\mu \nu})^2} \; \Big) , \nonumber \\
\label{LBI}
\end{eqnarray}
where $ \beta$ is an {\it a priori} unknown parameter with the dimension of [Mass]$^2$ that we write as $\beta \equiv M^2$,
and $\tilde{F}_{\mu \nu}$ is the dual of the field strength tensor ${F_{\mu \nu}}$.
Interest in Born-Infeld theory was revived in 1985 when Fradkin and Tseytlin~\cite{FT85} discovered that it appears
when an Abelian vector field in four dimensions is coupled to an open string, as occurs in models inspired by M theory 
in which particles are localized on lower-dimensional `branes' separated by a distance $\simeq 1/\sqrt{ \beta} = 1/M$ in some extra dimension~\footnote{Remarkably, 
the maximum field strength is related to the fact that the brane velocity is limited by the velocity of light~\cite{Bachas95}, confirming the insight of Born and Infeld~\cite{BornInfeld34}.}.
Depending on the specific brane scenario considered, $M$ might have any value between a few hundred GeV and the Planck scale $\sim 10^{19}$~GeV.
For the purposes of this paper, we consider only the relevant terms of fourth order in the gauge field strengths in (\ref{LBI}).

Until now, there has been no strong lower limit on the Born-Infeld scale $\beta$ or, equivalently,
the brane mass scale $M$ and the brane separation $1/M$. {A constraint corresponding to $M \gtrsim 100 $~MeV
was derived in~\cite{RSG} from electronic and
muonic atom spectra, though the derivation has been questioned in~\cite{CK}. Measurements of photon splitting in atomic fields~\cite{split}
were considered in~\cite{Davilaetal}, where it was concluded that they provided no limit on the Born-Infeld scale and it was suggested
that measurements of the surface magnetic field of neutron stars~\cite{neutronstar} might be sensitive to $M= \sqrt{ \beta} \sim 1.4 \times 10^{-5}$~GeV. 
More recently, measurements of nonlinearities in light by the PVLAS Collaboration~\cite{PVLAS14} are somewhat more sensitive to the individual nonlinear terms in
(\ref{LBI}), but are insensitive to the particular combination appearing in the Born-Infeld theory, as discussed in~\cite{Fouche16} where
more references can be found}. 

Here we show that the agreement of the recent
ATLAS measurement of light-by-light scattering with the standard QED prediction provides {the first limit on $M$ in the multi-GeV range, excluding a 
significant range extending to}
\begin{equation}
M \; \gtrsim \; 100~{\rm GeV} \, , 
\label{limit}
\end{equation}
entering the range of interest to brane theories. This limit is obtained under quite conservative assumptions, and plausible stronger assumptions
would strengthen our lower bound to $M \gtrsim 200$~GeV.

One may also consider a Born-Infeld extension of the Standard Model in which the hypercharge U(1)$_{\rm Y}$ gauge symmetry is realised non-linearly,
in which case the limit (\ref{limit}) is relaxed to
\begin{equation}
M_Y = \cos{\theta_W} M \gtrsim 90~{\rm GeV}, 
\label{Ylimit}
\end{equation}
where we have used $B_Y^\mu = {\rm cos}\theta_W\, A_{\rm EM}^\mu - {\rm sin}\theta_W \, Z^\mu$
and $\sin^2{\theta_W} \simeq 0.23$, with $\theta_W$ the weak mixing angle. 
As a corollary of this lower limit on the U(1)$_{\rm Y}$ brane scale, we recall that Arunasalam and Kobakhidze recently pointed out~\cite{ArunasalamKobakhidze17}
that the Standard Model modified by a Born-Infeld U(1)$_{\rm Y}$ theory has a finite-energy electroweak monopole~\cite{cho-maison,ChoKimYoon13} solution ${\cal M}$, 
whose mass they estimated as $M_{\cal M} \simeq 4~{\rm TeV} + 72.8 \, M_Y$. Such a monopole is less constrained by Higgs measurements than 
electroweak monopoles in other extensions of the Standard Model~\cite{EMY}, and hence of interest for potential detection by the ATLAS~\cite{atlasmono}, CMS and
MoEDAL experiments at the LHC~\cite{MoEDAL}. However, our
lower limit $M_Y \gtrsim 90$~GeV (\ref{limit}) corresponds to a 95\% CL lower limit on the mass of this monopole $M_{\cal M} \gtrsim 11$~TeV, 
excluding its production at the LHC.


\begin{figure}[h!]
\centering
\includegraphics[scale=0.5]{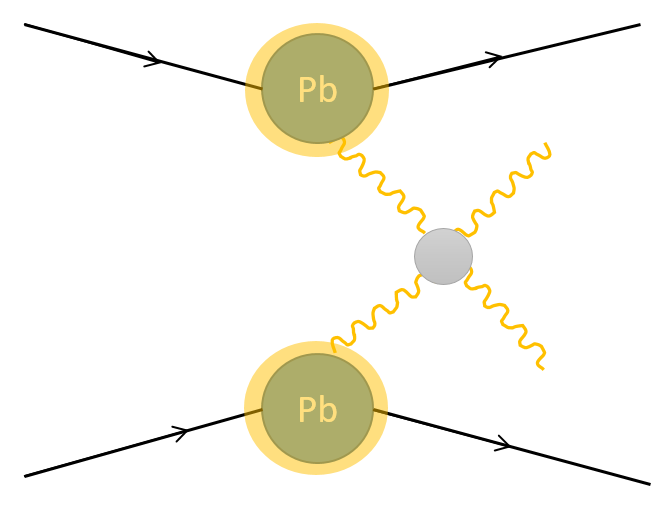}
\caption{\it Cartoon of light-by-light scattering through photon-photon collisions in ultra-peripheral Pb-Pb collisions. }
\label{fig:diagram}
\end{figure}

Following the suggestion of~\cite{dES13}~\footnote{These authors also considered the observability of light-by-light scattering in pp collisions at the LHC, but concluded that they were
less promising than Pb-Pb collisions.}, we consider ultra-peripheral heavy-ion collisions in which the nuclei scatter quasi-elastically via photon exchange: Pb + Pb ($\gamma \gamma$)$ \to $Pb$^{(*)}$ + Pb$^{(*)}$+ X, 
as depicted in Fig.~\ref{fig:diagram}, effectively acting via the equivalent photon approximation (EPA)~\cite{EPA} as a photon-photon collider. 
The EPA allows the electromagnetic field surrounding a highly-relativistic charged particle to be treated as equivalent to a flux of on-shell photons. 
Since the photon flux is proportional to $Z^2$ for each nucleus, the coherent enhancement in the exclusive $\gamma\gamma$ cross-section scales as $Z^4$, 
where $Z=82$ for the lead (Pb) ions used at the LHC. This is why heavy-ion collisions have an advantage over proton-proton or proton-lead collisions for 
probing physics in electromagnetic processes~\cite{dES13}. 
Photon fusion in ultra-peripheral heavy-ion collisions has been suggested as a way of detecting the Higgs boson~\cite{Higgsion, DEZ} and,
more recently, the possibility of constraining new physics beyond the Standard Model (BSM) in this process was studied in~\cite{Fichet, Knapenetal}.

\begin{figure}[h!]
\centering
\includegraphics[scale=0.4]{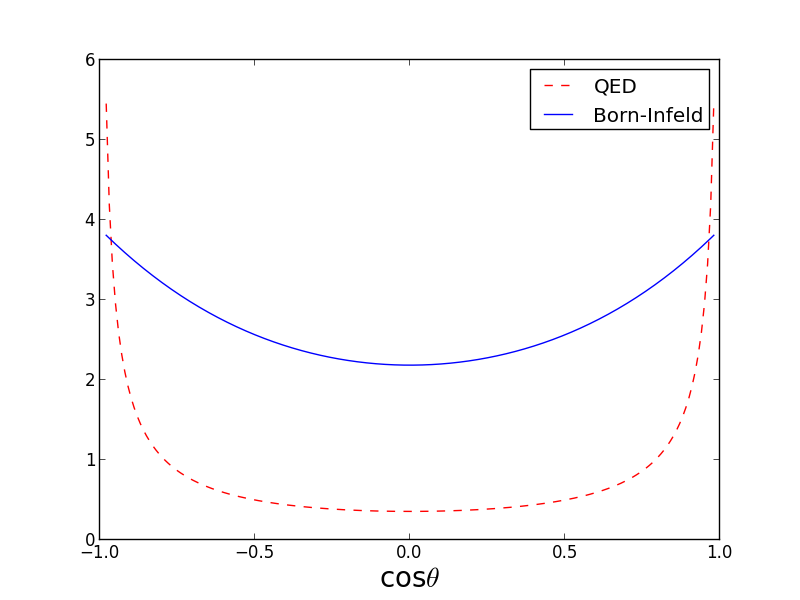}
\caption{\it Comparison between the angular distributions (with arbitrary normalisations) as functions of $\cos\theta$ in the centre-of-mass frame (where $\theta$ is the polar angle) for the
leading-order differential cross-sections in U(1)$_{\rm EM}$ Born-Infeld theory and QED, plotted as solid blue and dashed red lines, respectively.}
\label{fig:angulardistribution}
\end{figure}

As already mentioned, the possibility of directly observing light-by-light scattering at the LHC was proposed in~\cite{dES13},
and this long-standing prediction of QED was finally measured earlier this year with $4.4 \sigma$ significance by the ATLAS Collaboration~\cite{ATLAS17}
at a level in good agreement with calculations in~\cite{dES13, K-GLS16}. The compatibility with the Standard Model constrains any
possible contributions from BSM physics. Born-Infeld theory is particularly interesting in this regard,
as constraints from low-energy optical and atomic experiments have yet to reach the sensitivity of interest for measuring light-by-light scattering~\cite{PVLAS14, Fouche16}. 

The leading-order cross-section for unpolarised light-by-light scattering in Born-Infeld theory in the $\gamma \gamma$ centre-of-mass frame
 is given by~\cite{Davilaetal, TurkRebhan}:
\begin{equation}
\sigma_\text{BI}(\gamma\gamma \to \gamma\gamma) = \frac{1}{2}\int d\Omega \frac{d\sigma_\text{BI}}{d\Omega} = \frac{7}{1280\pi}\frac{m_{\gamma\gamma}^6}{\beta^4} \, ,
\label{BItotal}
\end{equation}
where $m_{\gamma\gamma}$ is the diphoton invariant mass and the differential cross-section is
\begin{equation}
\frac{d\sigma_\text{BI}}{d\Omega} =\frac{1}{4096\pi^2} \frac{m_{\gamma\gamma}^6}{\beta^4}\left(3 + \cos{\theta}\right)^2 \, .
\label{BIangle}
\end{equation}
We recall that the parameter $\beta = M^2$ enters as a dimensionful parameter in the Born-Infeld theory of non-linear QED defined by the Lagrangian (\ref{LBI}). If this originates from a Born-Infeld theory of hypercharge then the corresponding mass scale is $M_Y = \cos{\theta_W} M$. 

We plot in Fig.~\ref{fig:angulardistribution} the angular distributions as functions of $\cos\theta$ in the centre-of-mass frame (where $\theta$ is the polar angle) for the
leading-order differential cross-sections in both Born-Infeld theory and QED  (with arbitrary normalisations), as solid blue and dashed red lines, respectively. 
We see that the Born-Infeld distribution is less forward peaked than that for QED. For the latter, we used the leading-order amplitudes for the quark and lepton box 
loops in the ultra-relativistic limit from~\cite{Bernetal}, omitting the percent-level effects of higher-order QCD and QED corrections, as well the 
$W^\pm$ contribution that is negligible for typical diphoton centre-of-mass masses at the LHC. 

The total exclusive diphoton cross-section from Pb+Pb collisions is obtained by convoluting the $\gamma\gamma \to \gamma\gamma$ cross-section with a luminosity function $dL/d\tau$~\cite{CahnJackson},
\begin{equation}
\sigma_\text{excl.} = \int_{\tau_0}^1  d\tau \frac{dL}{d\tau} \sigma_{\gamma\gamma \to \gamma\gamma}(\tau) \, .
\end{equation} 
We have introduced here a dimensionless measure of the diphoton invariant mass, $\tau \equiv m_{\gamma\gamma}^2 / s_{NN}$, 
where $\sqrt{s_{NN}} = 5.02$ TeV is the centre-of-mass energy per nucleon pair in the ATLAS measurement. 
The luminosity function, derived for example in~\cite{CahnJackson}, can be written as an integral over the number distribution of photons carrying a fraction $x$ of the total Pb momentum:
\begin{equation}
\frac{dL}{d\tau} = \int_\tau^1 dx_1 dx_2 f(x_1) f(x_2) \delta(\tau - x_1 x_2) \, ,
\end{equation}
where the distribution function $f(x)$ depends on a nuclear form factor. We follow~\cite{CahnJackson} in adopting the form factor proposed in~\cite{DEZ},
while noting that variations in the choice leads to $\sim 20\%$ uncertainties in the final cross-sections~\cite{dES13}. A contribution with a non-factorisable 
distribution function should also be subtracted to account for the exclusion of nuclear overlaps, but this is not a significant effect
for the relevant kinematic range, causing a difference within the $20\% $ uncertainty~\cite{Knapenetal} from the photon luminosity 
evaluated numerically using the {\tt STARlight} code~\cite{starlight}. For $\sqrt{s_{NN}} = 5.5$ TeV and $m_{\gamma\gamma} > 5$ GeV 
we obtain a QED cross-section of $\sigma^\text{QED}_\text{excl.} = 385 \pm 77$ nb, in good agreement with~\cite{dES13}. 
The ATLAS measurement is performed at $\sqrt{s_{NN}} = 5.02$ TeV and for $m_{\gamma\gamma} > 6$ GeV,
for which we find $\sigma^\text{QED}_\text{excl.} = 220 \pm 44$ nb. 

This total $\gamma \gamma \to \gamma \gamma$ cross-section is reduced by the fiducial cuts of the ATLAS analysis,
which restrict the phase space to a photon pseudorapidity region $|\eta| < 2.4$, and require photon transverse energies $E_T > 3$ GeV 
and the diphoton system to have an invariant mass $m_{\gamma\gamma} > 6$ GeV with a transverse momentum $p_T^{\gamma\gamma} < 2$~GeV
and an acoplanarity $A\text{co} = 1 - \Delta\phi / \pi < 0.01$. We simulate the event selection using Monte-Carlo sampling, 
implementing the cuts with a 15\% Gaussian smearing in the photon transverse energy resolution at low energies and 0.7\% at higher energies~\cite{ATLAS17, ATLASphoton} 
above 100 GeV. 
Since the differential cross-section does not depend on $\phi$ we implement the acoplanarity cut as a fixed 85\% efficiency in the number of 
signal events after the $p_T^{\gamma\gamma}$ selection, following the ATLAS analysis~\cite{ATLAS17}. 
The total reduction in yield for the QED case is  a factor $\epsilon \sim 0.30$, which results in a fiducial cross-section
$\sigma^\text{QED}_\text{fid.} = 53 \pm 11$ nb for $\sqrt{s}_{NN} = 5.02$ TeV, in good agreement with the two predictions of 45 and 49 nb quoted by ATLAS~\cite{ATLAS17}. 

\begin{figure}[h!]
\centering
\includegraphics[scale=0.38]{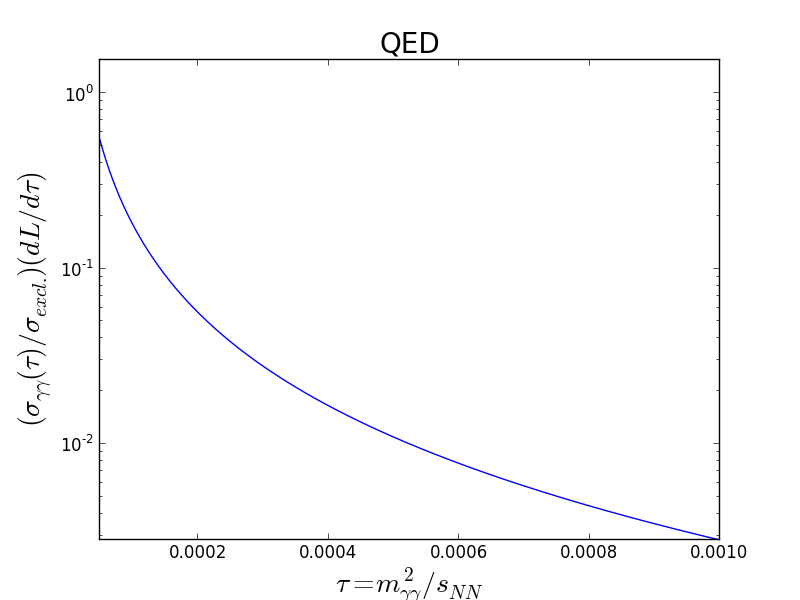}
\includegraphics[scale=0.38]{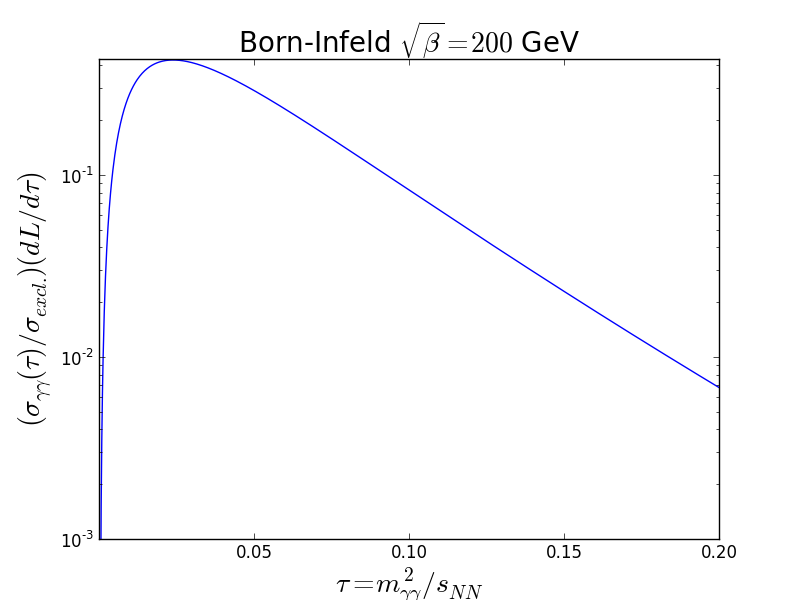}
\caption{ \it The distributions in the scaled diphoton invariant mass $\tau \equiv m_{\gamma\gamma}^2 / s_{NN}$,
normalised by the total $\gamma \gamma \to \gamma \gamma$ cross-section, 
for the QED case in the upper panel and for U(1)$_{\rm EM}$ Born-Infeld theory with $M = \sqrt{\beta} = 200$~GeV in the lower panel.}
\label{fig:invmdistribution}
\end{figure}

Following this validation for the QED case, we repeat the procedure for the Born-Infeld cross-section. Since the Born-Infeld $\gamma\gamma \to \gamma\gamma$ cross-section grows with energy, 
the dominant contribution to the cross-section comes from the $\tau \lesssim 0.2$  part of the integral, compared with $\tau \lesssim 10^{-4}$ for the QED case.
We show in Fig.~\ref{fig:invmdistribution} the distributions of the $\sigma(\gamma\gamma\to\gamma\gamma)$ cross-section
multiplied by the photon flux luminosity factor -- normalised by the total exclusive cross-section -- as functions of the invariant diphoton mass distribution, 
for the QED case in the left panel and in Born-Infeld theory with $M = \sqrt{\beta} = 200$ GeV in the right panel. 

We see that the invariant-mass distribution in the Born-Infeld case extends to $m_{\gamma \gamma} > M$, where the validity of the tree-level
Born-Infeld Lagrangian may be questioned because the Taylor expansion of the square root in the non-polynomial Born-Infeld Lagrangian (\ref{LBI}) could break down.
With this in mind, we use two approaches to place plausible limits on $M = \sqrt{\beta}$. In the first and most conservative method we consider 
$\gamma \gamma$ scattering only for $m_{\gamma \gamma} \le M$, while in the second approach we integrate the $\gamma \gamma$
cross-section (\ref{BItotal}) up to the diphoton invariant mass where the unitarity limit $\sigma_\text{BI} \sim 1/m_{\gamma\gamma}^2$ is saturated,
beyond which we assume that the cross-section saturates the unitarity limit and falls as $\sim 1/m_{\gamma\gamma}^2$. 

\begin{figure}[h!]
\centering
\includegraphics[scale=0.4]{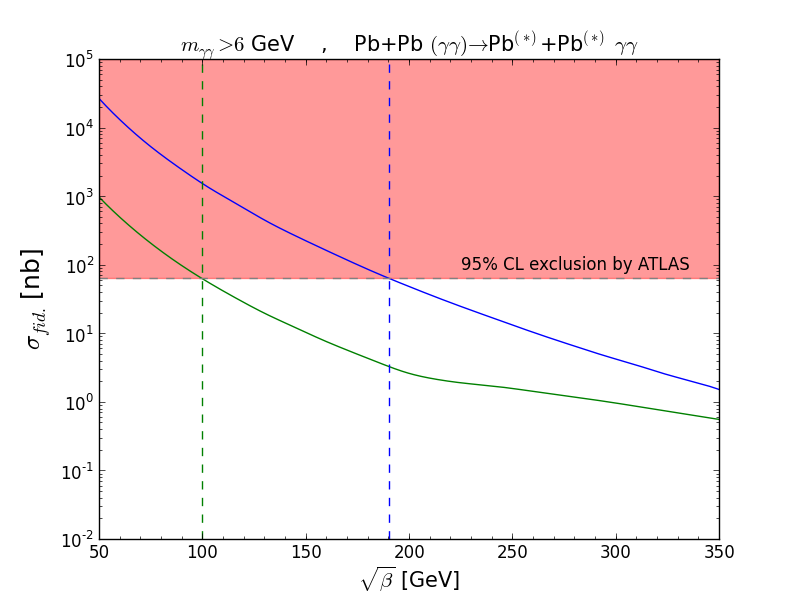}
\hspace{-0.8cm}
\includegraphics[scale=0.4]{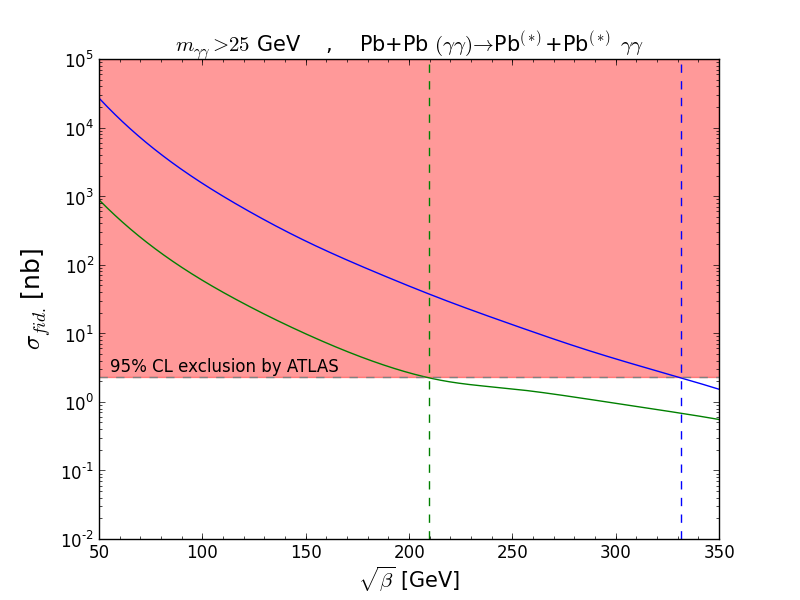}
\caption{\it The fiducial cross section for light-by-light scattering in relativistic heavy-ion collisions, $\sigma({\rm Pb} + {\rm Pb} (\gamma \gamma) \to
{\rm Pb}^{(*)} + {\rm Pb}^{(*)} \gamma \gamma)$ as a function of $M = \sqrt{\beta}$ in the U(1)$_{\rm EM}$ Born-Infeld theory is shown as a 
solid green (blue) line for a hard cut-off (unitarized) approach, respectively, as discussed in the text. The lower diphoton invariant mass cut-off is set at 6 GeV (25 GeV) on the upper (lower) plot.
This is compared with the 95\% CL upper limit obtained from the ATLAS measurement~\protect\cite{ATLAS17} by combining the statistical and systematic
errors in quadrature as well as a 10~nb theoretical uncertainty in the cross section predicted in QED~\protect\cite{dES13, K-GLS16} (horizontal dashed line),
which excludes the higher range shaded pink. The corresponding 95\% CL lower limits $M \gtrsim 100 \, (190)$~GeV for $m_{\gamma\gamma} > 6$ GeV, and $M \gtrsim 210 \, (330)$~GeV for $m_{\gamma\gamma} > 25$ GeV, are shown as vertical dashed lines in green (blue).}
\label{fig:betaconstraint}
\end{figure}

We find fiducial efficiencies for the cut-off and unitarization approaches to be $\epsilon \sim 0.39$ and $0.14$, respectively. 
Whilst  the $E_T$ and $\eta$ cuts have much less effect than for QED, as expected from the difference in the angular distributions visible in Fig.~\ref{fig:angulardistribution}, 
the larger invariant masses appearing in the Born-Infeld case are much more affected by the $p_T^{\gamma\gamma}$ requirement. 

Our calculations of the corresponding U(1)$_{\rm EM}$ Born-Infeld fiducial cross-sections are plotted in the left panel of Fig.~\ref{fig:betaconstraint} as a function of $M = \sqrt{\beta}$: the green curve is for
the more conservative cut-off approach, and the blue curve assumes that unitarity is saturated. These calculations are confronted with 
the ATLAS measurement of $\sigma_\text{fid.} = 70 \pm 24 \text{ (stat.)} \pm 17 \text{ (sys.)}$ nb~\cite{ATLAS17}, assuming that these errors are Gaussian 
and adding them in quadrature with a theory uncertainty of $\pm 10$ nb.
We perform a $\chi^2$ fit to obtain the 95\% CL upper limit on a Born-Infeld signal additional to the 49 nb Standard Model 
prediction~\footnote{We neglect possible interference effects that are expected to be small due to the different invariant mass and angular distributions involved.}.
This corresponds to the excluded range shaded in pink above $\sigma_\text{fid.}^{95\% \text{CL}} \sim 65$ nb in the left panel of Fig.~\ref{fig:betaconstraint}, 
which translates into the limit $M = \sqrt{\beta} \gtrsim 100 \, (190)$~GeV in the cut-off (unitarized) approach, 
as indicated by the green (blue) vertical dashed line in Fig.~\ref{fig:betaconstraint}, respectively.

These limits could be strengthened further by considering the $m_{\gamma \gamma}$ distribution shown in Fig.~3(b) of~\cite{ATLAS17}, 
where we see that all the observed events had $m_{\gamma \gamma} < 25$~GeV, in line with expectations in QED, whereas
in the Born-Infeld theory most events would have $m_{\gamma \gamma} > 25$~GeV. Calculating a ratio of the total exclusive cross-section of QED for $m_{\gamma\gamma} > 6$ GeV and $> 25$ GeV as $\sigma_\text{excl.}^{m_{\gamma\gamma}>25 \, \text{GeV}}/\sigma_\text{excl.}^{m_{\gamma\gamma}>6 \, \text{GeV}} \sim 0.02$, we estimate a 95\% CL upper limit of $\sim 2$ nb for $m_{\gamma\gamma} > 25$ GeV. The corresponding exclusion plot is shown in the right panel of Fig.~\ref{fig:betaconstraint}, where we see a stronger limit $M = \sqrt{\beta} \gtrsim 210 \, (330)$ GeV in the cut-off (unitarized) approach with the same colour coding as previously.

Our lower limit on the QED Born-Infeld scale $M = \sqrt{\beta} \gtrsim 100$~GeV is at least 3 orders of magnitude stronger than the
lower limits on $M = \sqrt{\beta}$ obtained from previous measurements of nonlinearities in light~\cite{RSG,CK,split,Davilaetal,PVLAS14,Fouche16}.
{Because of the kinematic cuts made in the ATLAS analysis, our limit does not apply to a range of values of $M \lesssim 10$~GeV
for which the nonlinearities in (\ref{LBI}) should be taken into account.
However, our limit is the first to} approach the range of potential interest for string/M theory constructions, since models with (stacks of)
branes separated by distances $1/M : M = {\cal O}(1)$~TeV have been proposed in that context~\cite{Tseytlin}. Our analysis could clearly be refined 
with more sophisticated detector simulations and the uncertainties reduced. However, in view of the strong power-law dependence of the Born-Infeld cross-section on $M = \sqrt{\beta}$ visible in
(\ref{BItotal}), the scope for significant improvement in our constraint is limited unless experiments can probe substantially larger $m_{\gamma \gamma}$
ranges. In this regard, it would be interesting to explore the sensitivities of high-energy $e^+ e^-$ machines considered as $\gamma \gamma$ colliders. 

As mentioned in the Introduction, Arunasalam and Kobakhidze have recently pointed out~\cite{ArunasalamKobakhidze17} that the Standard Model modified by a 
Born-Infeld theory of the hypercharge U(1)$_{\rm Y}$ contains a finite-energy monopole solution with mass $M_{\cal M} = E_0 + E_1$, where $E_0$ is the contribution associated with the
Born-Infeld U(1)$_{\rm Y}$ hypercharge, and $E_1$ is sssociated with the remainder of the Lagrangian.
Arunasalam and Kobakhidze have estimated~\cite{ArunasalamKobakhidze17} that $E_0 \simeq 72.8 \, M_Y$, where $M_Y = \cos{\theta_W} M$, and
Cho, Kim and Yoon had previously estimated~\cite{ChoKimYoon13} that $E_1 \simeq 4$~TeV~\footnote{Both these estimates are at the classical level,
and quantum corrections have yet to be explored.}. Combining these calculations and using our
lower limit $M \gtrsim 100$~GeV (\ref{limit}), we obtain a lower limit $M_{\cal M} \gtrsim 11$~TeV on the U(1)$_{\rm Y}$ Born-Infeld monopole mass~\footnote{For
completeness, we recall that it was argued in~\cite{ArunasalamKobakhidze17} that nucleosynthesis constraints on the abundance of relic monopoles
require $M_{\cal M} \lesssim 23$~TeV.}.
Unfortunately, this is beyond the reach of MoEDAL~\cite{MoEDAL} or any other experiment at the LHC~\cite{atlasmono}, but could lie within reach of a similar experiment at any future
100-TeV $pp$ collider~\cite{FCC-hh}, or of a cosmic ray experiment.

In this paper we have restricted our attention to possible Born-Infeld modifications of $U(1)$ gauge factors and their constraints in light-by-light scattering only. 
We plan to examine in the future the experimental constraints from measurements at the LHC on possible Born-Infeld extensions of the $SU(3)_C$ and $SU(2)_L$ gauge symmetries of the Standard Model.

\section*{Acknowledgements}

The work of JE and NEM was supported partly by the STFC Grant ST/L000326/1. 
The work of TY was supported by a Junior Research Fellowship from Gonville and Caius College, Cambridge.
We thank Vasiliki Mitsou for drawing our attention to~\cite{ArunasalamKobakhidze17},
and her and Albert De Roeck, Igor Ostrovskiy and Jim Pinfold of the MoEDAL Collaboration for their interest and relevant discussions. TY is grateful for the hospitality of King's College London where part of this work was completed.


\begin{thebibliography}{99}

\bibitem{Dirac28}
P.~A.~M.~Dirac,
  Proc.\ Roy.\ Soc.\ Lond.\ A {\bf 117} 610 (1928),
  doi:10.1098/rspa.1928.0023.
  
\bibitem{Dirac31}
  P.~A.~M.~Dirac,
  Proc.\ Roy.\ Soc.\ Lond.\ A {\bf 133} 60 (1931),
  doi:10.1098/rspa.1931.0130.

\bibitem{Halpern33}
O.~Halpern,
  Phys.\ Rev.\  {\bf 44} 855.2 (1993),
  doi:10.1103/PhysRev.44.855.2.
  
\bibitem{Heisenberg34}
W.~Heisenberg,
  Z.\ Phys.\  {\bf 90} (1934) 209
   Erratum: [Z.\ Phys.\  {\bf 92} 692 (1934)],
  doi:10.1007/BF01340782, 10.1007/BF01333516.

\bibitem{EulerKockel35}
H.~Euler and B.~Kockel,
  Naturwiss.\  {\bf 23} 246 (1935),
  doi:10.1007/BF01493898.
  
\bibitem{HeisenbergEuler36}
W.~Heisenberg and H.~Euler,
  Z.\ Phys.\  {\bf 98} 714 (1936),
  doi:10.1007/BF01343663
  [physics/0605038].
  
\bibitem{KarplusNeuman51}
R.~Karplus and M.~Neuman,
  Phys.\ Rev.\  {\bf 83} 776 (1951),
  doi:10.1103/PhysRev.83.776.
  
\bibitem{dES13}
D.~d'Enterria and G.~G.~da Silveira,
  Phys.\ Rev.\ Lett.\  {\bf 111}  080405 (2013),
   Erratum: [Phys.\ Rev.\ Lett.\  {\bf 116} 129901 (2016)]
  doi:10.1103/PhysRevLett.111.080405, 10.1103/PhysRevLett.116.129901
  [arXiv:1305.7142 [hep-ph]].
  
\bibitem{ATLAS17}
M.~Aaboud {\it et al.} [ATLAS Collaboration],
  arXiv:1702.01625 [hep-ex].
  
\bibitem{K-GLS16}
M.~K{\l}usek-Gawenda, P.~Lebiedowicz and A.~Szczurek,
  Phys.\ Rev.\ C {\bf 93} 044907 (2016)
  doi:10.1103/PhysRevC.93.044907,
  [arXiv:1601.07001 [nucl-th]].
  
\bibitem{BornInfeld34}
M.~Born and L.~Infeld,
  Proc.\ Roy.\ Soc.\ Lond.\ A {\bf 144} 425 (1994),
  doi:10.1098/rspa.1934.0059.
  
\bibitem{FT85}
E.~S.~Fradkin and A.~A.~Tseytlin,
  Phys.\ Lett.\  {\bf 163B} 123 (1985),
  doi:10.1016/0370-2693(85)90205-9.
  
\bibitem{Bachas95}
C.~Bachas,
  Phys.\ Lett.\ B {\bf 374} 37 (1996),
  doi:10.1016/0370-2693(96)00238-9
  [hep-th/9511043].
  
  \bibitem{RSG}
  J.~Rafelski, G.~Soff and W.~Greiner,
  Phys.\ Rev.\ A {\bf 7} (1973) 903.
  doi:10.1103/PhysRevA.7.903
  
  \bibitem{CK}
  H.~Carley and M.~K.-H.~Kiessling,
  Phys.\ Rev.\ Lett.\  {\bf 96} (2006) 030402
  doi:10.1103/PhysRevLett.96.030402
  [math-ph/0506069].
  
\bibitem{split}
  S.~Z.~Akhmadaliev {\it et al.},
  Phys.\ Rev.\ Lett.\  {\bf 89} (2002) 061802
  doi:10.1103/PhysRevLett.89.061802
  [hep-ex/0111084].
   
   \bibitem{Davilaetal}
J.~M.~D{\' a}vila, C.~Schubert and M.~A.~Trejo,
  Int.\ J.\  Mod.\ Phys.\ A\textbf{29} 1450174 (2014)
  doi:10.1142/S0217751X14501747
  [arXiv:1310.8410 [hep-ph]].
  
  
 \bibitem{neutronstar}
 S.~Mereghetti,
  Astron.\ Astrophys.\ Rev.\  {\bf 15} (2008) 225
  doi:10.1007/s00159-008-0011-z
  [arXiv:0804.0250 [astro-ph]].
  
  
\bibitem{PVLAS14}
F.~Della Valle {\it et al.},
  Phys.\ Rev.\ D {\bf 90} 092003 (2014),
  doi:10.1103/PhysRevD.90.092003.
  
\bibitem{Fouche16}
M.~Fouch{\' e}, R.~Battesti and C.~Rizzo,
  Phys.\ Rev.\ D {\bf 93}  093020 (2016), 
  doi:10.1103/PhysRevD.93.093020
  [arXiv:1605.04102 [physics.optics]].
  
\bibitem{ArunasalamKobakhidze17}
S.~Arunasalam and A.~Kobakhidze,
  arXiv:1702.04068 [hep-ph].
  
  
 \bibitem{cho-maison} Y.~M.~Cho and D.~Maison,
  Phys.\ Lett.\ B {\bf 391} 360 (1997)
  doi:10.1016/S0370-2693(96)01492-X
  [hep-th/9601028].

 
 
  \bibitem{ChoKimYoon13}
 Y.~M.~Cho, K.~Kim and J.~H.~Yoon,
  Eur.\ Phys.\ J.\ C {\bf 75} 67 (2015)
  [arXiv:1305.1699 [hep-ph]].
  
\bibitem{EMY}
J.~Ellis, N.~E.~Mavromatos and T.~You,
  Phys.\ Lett.\ B {\bf 756}  29 (2016)
  doi:10.1016/j.physletb.2016.02.048
  [arXiv:1602.01745 [hep-ph]].
  
  
  \bibitem{atlasmono} G.~Aad {\it et al.} [ATLAS Collaboration],
  Phys.\ Rev.\ Lett.\  {\bf 109}  261803 (2012)
  doi:10.1103/PhysRevLett.109.261803
  [arXiv:1207.6411 [hep-ex]];
  Phys.\ Rev.\ D {\bf 93}  no.5,  052009 (2016)
  doi:10.1103/PhysRevD.93.052009
  [arXiv:1509.08059 [hep-ex]].

  
  
  \bibitem{MoEDAL}
  B.~Acharya {\it et al.} [MoEDAL Collaboration],
  Int.\ J.\ Mod.\ Phys.\ A {\bf 29} 1430050 (2014),
  doi:10.1142/S0217751X14300506
  [arXiv:1405.7662 [hep-ph]];
  JHEP {\bf 1608}  067 (2016),
  doi:10.1007/JHEP08(2016)067
  [arXiv:1604.06645 [hep-ex]];
  Phys.\ Rev.\ Lett.\  {\bf 118} 061801 (2017),
  doi:10.1103/PhysRevLett.118.061801
  [arXiv:1611.06817 [hep-ex]].
  
  
   
  
\bibitem{EPA}
 C. von Weizsacker Z. Physik 88 612 (1934); E. J. Williams, Phys. Rev. 45  729 (1934); E. Fermi Nuovo Cimento 2  143 (1925).

\bibitem{Higgsion}
 E. Papageorgiu, 
 Phys. Rev. D{\bf 40}  92 (1989).
 
 
\bibitem{DEZ}
M.~Drees, J.~R.~Ellis and D.~Zeppenfeld,
  Phys.\ Lett.\ B {\bf 223}  454 (1989),
  doi:10.1016/0370-2693(89)91632-8.


\bibitem{Fichet}
S.~Fichet,
  arXiv:1609.01762 [hep-ph].
  
\bibitem{Knapenetal}
S.~Knapen, T.~Lin, H.~K.~Lou and T.~Melia,
  arXiv:1607.06083 [hep-ph].
  

  
\bibitem{TurkRebhan}
A.~Rebhan and G.~Turk,
  arXiv:1701.07375 [hep-ph].
  
\bibitem{Bernetal}
Z.~Bern, A.~De Freitas, L.~J.~Dixon, A.~Ghinculov and H.~L.~Wong,
  JHEP {\bf 0111} 031 (2001),
  doi:10.1088/1126-6708/2001/11/031
  [hep-ph/0109079].
  
\bibitem{CahnJackson}
R.~N.~Cahn and J.~D.~Jackson,
  Phys.\ Rev.\ D {\bf 42}  3690 (1990),
  doi:10.1103/PhysRevD.42.3690.


  
\bibitem{starlight}
S.~R.~Klein, J.~Nystrand, J.~Seger, Y.~Gorbunov and J.~Butterworth,
  Comput.\ Phys.\ Commun.\  {\bf 212}  258 (2017),
  doi:10.1016/j.cpc.2016.10.016
  [arXiv:1607.03838 [hep-ph]];
  {\tt https://starlight.hepforge.org}.
  
\bibitem{ATLASphoton}
G.~Aad {\it et al.} [ATLAS Collaboration],
  Eur.\ Phys.\ J.\ C {\bf 74}  no.10,  3071 (2014)
  doi:10.1140/epjc/s10052-014-3071-4
  [arXiv:1407.5063 [hep-ex]].

\bibitem{Tseytlin}
For a review and references, see A.~A.~Tseytlin,
  ``The many faces of the superworld", ed. M.~A.~Shifman, pp 417-452,
  doi:10.1142/9789812793850{\_}0025
  [hep-th/9908105].

  

  \bibitem{FCC-hh}
See, for example, the Future Circular Collider Study, \\
{\tt https://fcc.web.cern.ch/Pages/default.aspx}.

 \end{thebibliography}
\end{document}